\def \yskip{\penalty-50\vskip3pt plus 3pt minus 2pt}
\def \reference{\par \yskip \noindent \hangindent .4in \hangafter 1}
\def \abc#1#2#3#4 {\reference#1, {\sl#2}, {\bf#3}, #4}
\def \blank {\lower 5pt\hbox to 0.75in{\hrulefill}}

\def \s{~\rm{s}}
\def \km{~\rm{km}}

\def \erg{~\rm{erg}}

\def \lesssim{\mathrel{<\kern-1.0em\lower0.9ex\hbox{$\sim$}}}
\def \gtrsim{\mathrel{>\kern-1.0em\lower0.9ex\hbox{$\sim$}}}

\documentstyle[11pt,aasms,tighten,flushrt]{article}
\begin{document}
%\normalsize
\small

\setcounter{page}{1}

%\shorttitle{COMMON ENVELOPE EVOLUTION}
%\shortauthors{SOKER}

%\slugcomment{Draft version of \today}

\title{COMMENTS ON THE FINAL ORBITAL SEPARATION IN
COMMON ENVELOPE EVOLUTION}

\author{Noam Soker}

\affil{Department of Physics, Oranim, Tivon 36006, Israel; 
soker@physics.technion.ac.il}
                                    
%\altaffiltext{1}{On sabbatical from the University of Haifa at Oranim,
%Department of Physics, Oranim, Tivon 36006, Israel}

$$
$$

\centerline {\bf ABSTRACT}

I study some aspects of common envelope evolution, where a compact
star enters the envelope of a giant star.
I show that in some binary systems under a narrow range of
parameters, a substantial fraction of the giant stellar envelope
is lost before the onset of the common envelope.
The reduced envelope mass at the onset of the common envelope
implies that the binary system emerges from the common envelope
with a relatively large orbital separation.
I therefore caution against a simple treatment, which omits
this process, in the study of systems that eveloved through
a common envelope phase and ended with
a relatively large orital separation, e.g., PG1115+166 .
A fraction of the envelope that is lost while the companion
is still outside the giant stellar envelope is accreted
by the companion. The companion may form an accretion disk
and blow two jets.
I propose this scenario for the formation of the bipolar
planetary nebula NGC 2346, which has a binary nucleus with an
orbital period longer than that of any other known binary system
in planetary nebulae.
Because this scenario, of a late common envelope phase,
occurs for a narrow range of parameters, I expect it to be
applicable to a small, but non-negligible, number of systems.

{\it Subject headings:} stars: binaries
$-$ stars: evolution
$-$ stars: AGB and post-AGB
$-$ stars: mass loss

% ===================================================
\section{INTRODUCTION} 
% ===================================================

 Many evolved close binary systems acquire their close orbit
through common envelope evolution (Paczynski 1971; Iben \& Livio
1993; Armitage \& Livio 2000; Taam \& Sandquist 2000).
In the common envelope phase a compact companion enters
the envelope of a more extended star, e.g., an asymptotic giant
branch (AGB) star, and because of tidal interaction and friction,
the orbit shrinks (see review by Iben \& Livio 1993). 
In principle, orbital energy and angular momentum are transferred
to the giant stellar envelope, where part of the orbital energy
goes to expel the envelope.
 A commonly used parameter is the ratio of the orbital energy
that is released during the common envelope phase,
$\Delta E_{\rm orb}$, to the binding energy of the ejected envelope,
$\Delta E_{\rm bind}$:
$\alpha_{\rm CE} \equiv \Delta E_{\rm bind}/\Delta E_{\rm orb}$
(e.g., Livio \& Soker 1988).
Note that different definitions for the
binding energy exist (e.g., O'Brien, Bond, \& Sion 2001).
Since the orbital energy that is released depends mostly on the final
orbital separation, the value of $\alpha_{\rm CE}$ can in principle
be calculated for systems whose final orbital separation is known.
This value can then be used to derive the final separation in the
study of the common envelope phase in binary systems evolution.
To calculate the value of $\alpha_{\rm CE}$ in a specific system,
the giant structure at the onset of the common envelope should be
known, in particular the envelope mass and radius.
Using different values for the envelope mass and radius
yields a very different value for $\alpha_{\rm CE}$
(O'Brien et al.\ 2001; Maxted et al.\ 2002).

Maxted et al.\ (2002) assume that negligible mass has been lost
prior to the onset of the common envelope in PG1115+166, and
find $\alpha_{\rm CE} >1$, so they propose,
that internal energy stored in the envelope,
e.g., ionized gas, contributes to envelope ejection.
I disagree with this proposal for the following reasons.
First, Harpaz (1998) convincingly argues that the energy
released in the recombination of the envelope does not contribute
much to the ejection of the envelope.
Second, on a more general ground, energy is not a problem in
the mass loss process of AGB stars.
In most AGB stars, the momentum carried by the wind, $\dot M_w v_w$,
is not much above the momentum supplied by the radiation $L/c$,
where $\dot M_w$ is the mass loss rate into the wind, $v_w$
the wind speed, $c$ the speed of light, and $L$ the stellar luminosity.
In most systems $\dot M_e v_w \lesssim 3 L/c$ (Knapp 1986),
implying that the ratio of stellar luminosity to the
rate of kinetic energy carried by the wind is
$L/\dot E_w \gtrsim 10^4 $, assuming $v_w \sim 20 \km \s^{-1}$.
Thus, the energy is not a problem as regards the ejection of an AGB
stellar envelope, and it is not clear how
relevant the $\alpha_{\rm EC}$ parameter is by itself.
This can be demonstrated for PG1115+166, for which 
the final masses are $0.43 M_\odot$ and $0.52 M_\odot$, and
the final orbital separation is $a_f = 45 R_\odot$
(Maxted et al. 2002). The orbital energy that was released
during the spiraling-in process was
$\Delta E_{\rm orb} \simeq 10^{46} \erg$.
An AGB star with a luminosity of $5000 L_\odot$ releases this
energy in only $< 20$ years.
 
In any case, another way to reduce the calculated value of
$\alpha_{\rm CE}$ is to reduce the binding energy of the
envelope by considering a decrease of its mass before
the onset of the common envelope.
This is the goal of the present paper.
I show that for a small initial parameters space,
the companion brings the giant envelope to corotate with the
orbital motion, hence the spiraling-in process stops before the
companion reaches the envelope, such that only after the
giant star loses a substantial fraction of its envelope
does the compact companion enter the envelope.
In $\S 2$ I discuss the initial stage of reaching synchronization,
and in $\S 3$ I discuss the onset of the common envelope.
The general evolution of a close companion to a giant star was treated
before, e.g., Alexander, Chau, \& Henriksen (1976), who include
accretion by the companion, drag, tidal force, and eccentricity.
They note that the effect of a wind blown by a rotating giant
is to reduce the orbital separations, but do not present
a detailed solution for the orbital separation.
In the present paper I neglect some effects studied in the past.
This allows a simple solution for the orbital separation,
which clearly demonstrates the points raised here. 
I discuss this solution and its application to specific systems
in $\S 3$. My short summary is in $\S 4$.

% ===================================================
\section{REACHING SYNCHRONIZATION} 
% ===================================================

I treat systems in which the secondary compact star,
i.e., a main sequence star, a white dwarf, or a neutron star,
is massive enough to bring the mass-losing giant star to
synchronization.
I neglect eccentricity, mass accretion and accretion drag
by the companion, effects which were studied in the past
(e.g., Alexander et al.\ 1976). 
I do consider tidal interaction, assuming synchronization is reached,
and the effect of mass loss from a rotating giant. 
The system is taken to be stable to the Darwin instability.
This stability criterion is 
$I_{\rm orb} > 3 I_{\rm env}$ (e.g., Eggleton \& Kiseleva-Eggleton 2001),
where $I_{\rm orb}$ is the moment of inertia of the binary system,
$I_{\rm env} = \eta M_{\rm env} R_g^2$ is the moment
of inertia of the giant, $R_g$ is the giant's radius,
$M_{\rm env}$ is the envelope mass, and $\eta \simeq0.2$
for AGB stars.
After substituting for the orbital moment of inertia,
the criterion for Darwin stability reads
\begin{equation}
\frac {M_g M_ 2}{M_g + M_2} a^2 \gtrsim 3 \eta M_{\rm env} R_g^2,
\end{equation}
where $a$ is the orbital separation, and $M_g$ the giant stellar mass.
Because of tidal interaction, orbital angular momentum is transferred
to the giant's envelope, and the orbit starts to shrink  
when the giant's radius $R_g$ increases to $R_g \sim 0.1-0.2 a$.
The exact ratio $R_g/a$ when synchronization and circularization occur
strongly depend on the eccentricity (Hut 1982;
Eggleton \& Kiseleva-Eggleton 2001), and weakly on some of the giant's
properties, e.g., evolution time (see scaled equations in Soker 1996).
 After the system reaches synchronization, and the spiral-in process
stops for a while, the giant radius is likely to further increase.
I take an orbit with initial semi-major axis of $a_i$,
while the radius when synchronization and circularization are achieved
is $a_0<a_i$.
To derive a simple expression, I take $M_2 \ll M_g$ at this
evolutionary phase, before the giant loses most of its mass.
 The orbital angular momentum released in the spiraling-in
from $a_i$ to $a_0$ is
$\Delta J_{\rm orb} \simeq M_2 (G M_g)^{1/2}(a_i^{1/2}-a_0^{1/2})$.
This equals the angular momentum gained by the envelope, which,
assuming that initially the envelope has no angular momentum, is
$J_{\rm env}=I_{\rm env} \omega_0$, where
$\omega_0= (G M_g/a_0^3)^{1/2}$ is the angular velocity of the envelope
at the onset of synchronization.
 The equality $\Delta J_{\rm orb}=J_{\rm env}$ gives
\begin{equation}
\left[ \left( \frac{a_i}{a_0} \right)^{1/2} - 1 \right]
\left( \frac{a_0}{R_g} \right)^2 \simeq \frac {0.2 M_{\rm env}}{M_2}.
\end{equation}

Strong tidal interaction starts at $a_0 \sim 5-7 R_g$ (Soker 1996).
For $M_2 = 0.1 M_{\rm env}$ and $a_i=7R_g$, synchronization is
reached at $a_0 \sim 6 R_g$ by equation (2).
The system is still stable to the Darwin instability (eq. 1).
Much lower mass secondary stars will spiral-in all the way to the
envelope, because $a_0$ will be smaller and the orbit becomes
unstable by equation (1).
 For the envelope to influence the orbital motion, its mass can't
be much below that of the companion.
Assuming that most of the initial giant mass $M_g$ is in the envelope,
I find that the arguments of this paper are applicable to secondary stars
in the mass range
\begin{equation}
0.1 M_{g0} \lesssim M_2 \lesssim 0.5 M_{g0},
\end{equation}
where $M_{g0}$ is the initial giant mass (when synchronization
occurs).

% ===================================================
\section{ENTERING A COMMON ENVELOPE PHASE} 
% ===================================================

Assuming again a synchronized circular orbit, the angular velocity
of both the orbital motion and the giant's envelope is
\begin{equation}
\omega=\left[ \frac {G(M_2+M_g)}{a^3} \right]^{1/2}.
\end{equation}
 The wind blown by the giant carries angular momentum at a rate
given by 
\begin{equation}
\dot J_{\rm wind} = (a_g^2+\beta R_g^2) \omega \dot M_w, 
\end{equation}
where a dot stands for a time derivative, $\dot M_w = -\dot M_g$ is the
rate of mass blown in the wind,
$a_g=a M_2/(M_g+M_2)$ is the distance of the mass-losing giant
star from the center of mass, and $\beta$ is a factor which depends
on the mass-loss geometry: $\beta=2/3$ for a uniform mass loss from
the giant's surface, while $\beta=1$ for a mass loss from the
equatorial plane.
The first term inside the parenthesis of equation (5) is the
angular momentum of the wind due to the motion around the center
of mass, while the second term is due to the rotation of the envelope.
The sources of the angular momentum carried by the wind are the
orbital and giant envelope's angular momentum.
 The orbital angular momentum changes at a rate of
\begin{equation}
\frac{ \dot J_{\rm orb}}{J_{\rm orb}} =
\frac {\dot M_g}{M_g} - \frac{\dot M_g}{2(M_g+M_2)}
+ \frac {\dot a}{2a}, 
\end{equation}
where $J_{\rm orb}$ is the orbital angular momentum, and, as
earlier, I assume a circular orbit.
Neglecting the change in $R_g$ and in the core mass,
$M_{\rm core} = M_g-M_{\rm env}$, the rate of change of the
giant's envelope angular momentum is
\begin{equation}
\dot J_{\rm env} = \frac{1}{2} \omega \eta M_{\rm env} R_g^2 
\left( 2 \frac {\dot M_g}{M_{\rm env}} + \frac {\dot M_g}{M_g+M_2}
-3 \frac{\dot a}{a} \right),
\end{equation}
where $\eta$ is defined before equation (1).
 Conservation of angular momentum implies the equality
$\dot J_{\rm orb}+ \dot J_{\rm env}+\dot J_w =0$.
Using equations (5)-(7), the last equality gives the expression
for the rate of change of orbital separation $\dot a$ as
\begin{eqnarray}
\frac {\dot a}{a} \left[ 1 - 3 \eta M_{\rm env} 
\frac {M_g+M_2}{M_g M_2} \left( \frac {R_g}{a} \right)^2 \right] 
= - \frac {\dot M_g}{M_g+M_2}  \nonumber \\
\times \left[ 1 -
2 \beta \frac {(M_g+M_2)^2}{M_g M_2} \left( \frac {R_g}{a} \right)^2
+ 2 \eta \frac {(M_g+M_2)^2}{M_g M_2} \left( \frac {R_g}{a} \right)^2
+ \eta M_{\rm env} \frac {M_g+M_2}{M_g M_2} \left( \frac {R_g}{a} \right)^2
\right].
\end{eqnarray}
For a non-rotating giant and/or $R_g \ll a$, the familiar expression
$(\dot a/a)=- \dot M_g/(M_g+M_2)$ is recovered, and the orbit expands
with mass loss.
For close orbits, where both $R_g$ is not much smaller than $a$
and the giant rotates fast, the second term in the
square brackets on the right hand side dominates,
and the orbit shrinks.
The binary system may enter a common envelope phase, but only
after some fraction of the envelope mass has been lost.

To cast equation (8) in a simple form, I define
\begin{equation}
\chi \equiv  1 - 3 \eta M_{\rm env} 
\frac {M_g+M_2}{M_g M_2} \left( \frac {R_g}{a} \right)^2 .
\end{equation}
The stability against Darwin instability, equation (1), is
$\chi > 0$.
I also define
\begin{equation}
\mu \equiv \frac {M_g}{M_2}.
\end{equation}
Because the accretion by the companion is neglected,
$\dot M_g = M_2 \dot \mu$. 
 Canceling the time derivative, equation (8) can be written in the
form 
\begin{equation}
\chi \frac {d a^2}{d \mu} = 
- \frac {2 a^2}{\mu+1}
+ 4 R_g^2 \frac {\beta (\mu + 1)}{\mu}
- 4 \eta R_g^2 \frac {\mu + 1}{\mu} 
- 2 \eta R_g^2 \frac {M_{\rm env}}{M_g}. 
 \end{equation}
For the effects studied here to be important,
$2 \lesssim \mu \lesssim 10$ (eq. 3).
Taking this range for the value of $\mu$, and the
range for the the value of the wind geometrical factor
$(2/3) \leq \beta \leq 1$, I approximate $\beta (\mu+1)/\mu \simeq 1$.
In light of the effects neglected here, e.g., accretion by the
companion, I also make the approximations
that $\chi$ is constant, $\mu+1/\mu \simeq 1$, and
$M_{\rm env}/M_g \simeq 1$. 
These assumptions yield a simple form for equation (11):
\begin{equation}
\chi \frac {d a^2}{d \mu}  \simeq
- \frac {2 a^2}{\mu+1} + (4-6\eta) R_g^2,
\end{equation}
which has a simple solution
\begin{equation}
\left( \frac {a}{R_g} \right)^2 \simeq
\left( \frac {a_0}{R_g} \right)^2
\left( \frac {\mu_0+1}{\mu+1} \right)^{2/\chi}
-\frac{4-6\eta}{2+\chi}
(\mu_0+1) \left( \frac {\mu_0+1}{\mu+1} \right)^{2/\chi}
+\frac{4-6\eta}{2+\chi} (\mu+1), 
\end{equation}
where the initial condition, after synchronization and circularization
are reached via strong tidal interaction, is $a(\mu_0)=a_0$.
 
% ===================================================
\section{DISCUSSION} 
% ===================================================

When $R_g \ll a_0$ the first term on the right hand side of
equation (13) dominates,  $\chi=1$, and the familiar
expression for orbit expansion with mass loss is recovered.
For a general case, when condition (1) holds and $0 < \chi <1$,
then the factor $[(\mu_0+1)/(\mu+1)]^{2/\chi}$ that multiplies
the first two terms on the right hand side of equation (13)
becomes very large compared with the third term.
In that case, there are three cases which are determined by the
ratio of the two first terms on the right hand side
\begin{equation}
\Gamma \equiv \left( \frac {R_g}{a_0} \right)^2 
\frac{4-6\eta}{2+\chi} (\mu_0 + 1).
\end{equation}

If $\Gamma=1$, the first two terms cancel each other,
and the orbit shrinks according to $a_0 \simeq R_g(\mu+1)^{1/2}$,
where I assumed $(4-6\eta)/(2+\chi) \simeq 1$.
 A common envelope will occur only if the giant radius increases
a little during this process, and $\Gamma$ becomes $>1$. 
In the first case, occurring for $\Gamma \leq 1$,
there will be no common envelope phase.
In the second case, for $\Gamma \gg 1$ (as I show below,
practically for $\Gamma \gtrsim 1.3$), the orbit decreases
fast with mass loss, and a common envelope starts after only
a small amount of the envelope mass has been lost.

In the third case, the one this paper aims at, the onset of the
common envelope occurs after a large fraction of the envelope
mass has been lost.
Consider the following case, which I choose to demonstrate the third
case.
I take $a_0/R_g=2.5$, $\eta=0.2$, $\mu_0=M_g/M_2=6$,
and an average, over time, value of $\chi=0.5$. 
The value of $\mu_0$ was chosen to match that of the
progenitor of PG1115+166 (Maxted et al.\ 2002). 
These values give $\Gamma=1.25$.  
The orbit shrinks, and the common envelope phase starts, i.e., $a=R_g$,
when $\mu \simeq 4.3$.
 If the giant's core mass is $M_{\rm core} \simeq M_2$,
these values of $\mu_0=6$ and $\mu=4.3$ mean that the envelope
mass decreased to $\sim 2/3$ its initial mass.
 For higher values of $\Gamma$, the common envelope phase will start
at earlier stages (depending on the other parameters as well),
with a small decrease in envelope mass.
I conclude that the scenario I am aiming at, i.e., that
$\gtrsim 50 \%$ of the envelope is lost before the onset of the
common envelope, is applicable for a narrow
range of parameters, for which $1<\Gamma \lesssim 1.3$.
 Because $\chi$, and to lesser degree $R_g$, change during
the evolution, the exact range for $\Gamma$ should be determined
by numerical evolutionary calculations.
I propose that the progenitor of PG1115+166 had parameters similar to
the values chosen above, and in consequence a large fraction of the
envelope mass was lost before the onset of the common envelope.
The low envelope mass at the onset of the common envelope means a low
binding energy. This, I suggest, accounts for the relatively large
orbital separation found by Maxted et al.\ (2002).

 The evolutionary scenario studied here is relevant also to the
formation of some bipolar planetary nebulae (PNs).
Bipolar PNs are PNs with two lobes and an equatorial
waist between them are similar in many respects to
bipolar nebulae around symbiotic systems (Corradi \& Schwarz 1995;
Corradi 1995), and are thought to be formed from binary systems.
In most cases the binary companion to the AGB star is outside the
envelope (Soker \& Rappaport 2000).
 From the 16 PNs known to have close-binary nuclei (Bond 2000),
only NGC 2346 is a bipolar PNs.
 The orbital period of the central-binary system in NGC 2346 is
$P_{\rm orb} = 16~$day, much longer than those of the other 12
PNs with determined orbital periods, all having $P_{\rm orb}< 3~$day.
It is likely, therefore, that the companion to the AGB progenitor
of NGC 2346 entered the envelope when the envelope mass
was already small, so it did not spiral-in as close to the
AGB core as in the other systems.
I propose that during most of the evolution time the companion was
outside, but close to, the envelope, where it accreted mass and
blown collimated fast wind (CFW), or jets, which shaped the bipolar
lobes (Soker \& Rappaport 2000).
Because of the narrow range required for the above scenario,
$1< \Gamma \lesssim 1.3$, I do not expect to find many
bipolar PNs with close binary systems at their centers.
                                      
% ===================================================
\section{SUMMARY} 
% ===================================================

The main purpose of this paper is to caution against a simple
use of the $\alpha_{\rm CE}$ parameter when studying the final
separation of binary systems emerging from a common envelope.
$\alpha_{\rm CE}$ is defined as the ratio of the
binding energy of the envelope to the orbital energy
released by the spiraling-in binary system.

There is no need to invoke an extra energy source if
a simple calculation shows that in a specific system
$\alpha_{\rm CE} > 1 $.
First, plenty of energy is supplied by the nuclear burning in
the giant's core, $\gtrsim 10^4$ times the rate of kinetic energy
carried by the wind.
 The gravitational energy released by the companion spiraling-in inside
the giant envelope, though, has the advantage of being
transferred in part directly into mechanical energy, by spinning-up
the envelope, by tidal interaction, and by exciting pressure modes.
Possible jets blown by the companion also have this advantage
(Armitage \& Livio 2000). 
(A small fraction of the nuclear energy is also transferred
to mechanical energy via stellar pulsation,
which becomes very strong on the upper AGB.)
 
Second, as I showed in this paper,
in some systems a substantial fraction of the envelope
may be lost before the onset of the common envelope, 
thereby reducing the envelope's binding energy,
hence the value of the calculated $\alpha_{\rm CE}$.
 The evolution is as follows.
 As the giant radius increases, tidal interaction becomes stronger.
When the ratio of the giant radius to the orbital separation
becomes $R_g/a \gtrsim 0.1-0.2$, and the giant mass to companion
mass ratio is $\mu \equiv M_g/M_2 \lesssim 10$, the system reaches
synchronization and circularization.
The spiraling-in process stops. 
As mass loss proceeds, angular momentum is lost from the system.
The giant most likely continues to expand and the ratio
$a/R_g$ decreases, so the effective value of $a_0/R_g$ becomes
smaller than that at the moment synchronization is achieved.
The condition for this scenario to occur is
$1 < \Gamma \lesssim 1.3$, where $\Gamma$ is given in equation (14).
If $\Gamma \leq 1$ the system does not enter a common envelope phase,
while if $\Gamma \gtrsim 1.3$ the system enters a common envelope
phase after only a small amount of envelope has been lost.
The exact condition of $\Gamma$ depends on the values of the
different parameters, as well as on some effects not studied here,
e.g., accretion by the companion, drag, and eccentricity. 

This scenario is applicable to systems which went through a common
envelope and end with a relatively large orbital separation,
e.g., PG1115+166 (Maxted et al.\ 2002).
 When outside the envelope, the companion can accrete from the
AGB wind, form an accretion disk and blow jets.
This will form a bipolar planetary nebula.
 Later the companion enters the envelope and spirals-in, ending
at closer orbit to the core. I proposed this scenario to
account for the bipolar planetary nebula NGC 2346, which has a
binary nucleus with orbital period of 16 days (Bond 2000).
Because this scenario occurs for a narrow range of parameters,
I expect it to be applicable to a small, but non-negligible,
number of systems.
                                                                   
\acknowledgements

This research was supported by the US-Israel Binational Science Foundation.


\begin{references}

\reference{} Alexander, M. E., Chau, W. Y., \& Henriksen, R. N.
   1976, ApJ, 204, 879

\reference{} Armitage, P. J., \& Livio, M. 2000, ApJ, 532, 540

\reference{} Bond, H. 2000, in Asymmetrical Planetary Nebulae II:
     From Origins to Microstructures,
    Astronomical Society of the Pacific Conference Series, Vol. 199,
    eds. J. Kastner, N. Soker, \& S. Rappaport, 115

\reference{} Corradi, R. L. M. 1995, MNRAS, 276, 521 

\reference{} Corradi, R. L. M., \& Schwarz, H. E. 1995, A\&A, 293, 871 

\reference{} Eggleton, P. P., \& Kiseleva-Eggleton, L. 2001,
   ApJ, 562, 1012 

\reference{} Harpaz, A. 1998, ApJ, 498, 293 

\reference{} Hut, P. 1982, A\&A, 110, 37

\reference{} Iben, I. Jr., \& Livio, M. 1993, PASP, 105, 1373 

\reference{} Knapp, G. R. 1986, ApJ, 311, 731

\reference{} Livio, M., \& Soker, N. 1988, ApJ, 329, 764

\reference{} Maxted, P. F. L., Burleigh, M. R., Marsh, T. R.,
      \& Bannister, N. P. 2002, MNRAS, in press (astro-ph/0203519)

\reference{} O'Brien, M. S., Bond, H. E., \& Sion, E. M. 2001,
            ApJ, 563, 971 

\reference{} Paczynski, B. 1971, ARA\&A, 9, 183

\reference{} Soker, N. 1996, ApJ, 460, L53

\reference{} Soker, N., \& Rappaport, S. 2000, ApJ, 538, 241 

\reference{} Taam, R. E., \& Sandquist, E. L. 2000, ARA\&A, 38, 113

\end{references}
\end{document}